\documentclass[notitlepage,10pt,aps,pre,showpacs,superscriptaddress,groupedaddress,amsmath,amssymb]{revtex4-1}
\usepackage[normalem]{ulem}
\usepackage{epsfig}
\usepackage{graphics}
\usepackage{lipsum}
\usepackage{sidecap}
\usepackage{amsmath} 
\usepackage{mathtools}
\usepackage{graphicx} 
\usepackage{bm}
\usepackage{xcolor,colortbl}
\usepackage{color, colortbl}
\definecolor{darkgray}{rgb}{0.66, 0.66, 0.66}
\definecolor{yellow-green}{rgb}{0.6, 0.8, 0.2}
\definecolor{deeppink}{rgb}{1.0, 0.08, 0.58}
\definecolor{darkviolet}{rgb}{0.58, 0.0, 0.83}

\definecolor{darkcyan}{rgb}{0.0, 0.55, 0.55}
\begin{document}
\title{Spin Torque Oscillations Triggered by In-plane Field}
\author{R. Arun$^{1}$,R. Gopal$^{2}$, V.~K.~Chandrasekar$^{2}$, and M.~Lakshmanan$^1$,}
\address
{
$^{1}$Department of Nonlinear Dynamics, School of Physics, Bharathidasan University, Tiruchirapalli-620024, India\\
$^{2}$Centre for Nonlinear Science \& Engineering, School of Electrical \& Electronics Engineering, SASTRA Deemed University, Thanjavur- 613 401, India. \\
}

\date{\today}
	
\begin{abstract}
We study the dynamics of a spin torque nano oscillator that consists of parallelly magnetized free and pinned layers by numerically solving the associated Landau-Lifshitz-Gilbert-Slonczewski equation in the presence of a field-like torque.  We observe that an in-plane magnetic field which is applied for a short interval of time ($<$1ns) triggers the magnetization to exhibit self-oscillations from low energy initial magnetization state.  Also, we confirm that the frequency of oscillations can be tuned over the range $\sim$25 GHz to $\sim$72 GHz by current, even in the absence of field-like torque.  We find the frequency enhancement up to 10 GHz by the presence of field-like torque. We determine the Q-factor for different frequencies and show that it increases with frequency. Our analysis with thermal noise confirms that the system is stable against thermal noise and the dynamics is not altered appreciably by it.
\end{abstract}

\vspace{2pc}


\maketitle

\section{Introduction}
The spin torque nano oscillator (STNO) is a well-known candidate for its potential application for microwave generation in the GHz range~\cite{Grollier,Slavin,Locatelli}.  Basically, the nanosized STNO consists of two ferromagnetic layers, a free layer, and pinned layer, where the former one can change its direction of magnetization while the latter one corresponds to a fixed magnetization. The free layer is comparatively thinner than the pinned layer.  These two layers are separated by a nonmagnetic but conductive or insulative layer called a spacer.  The magnetization of the free layer can be made to oscillate by tuning the current and/or magnetic field and the magnetization oscillations can be converted into voltage oscillations by the Giant Magnetoresistive effect~\cite{Slon,Berger,Kiselev,Rippard}.  

In an STNO the magnetization of the free layer is manipulated by the transfer of spin angular momentum between the incoming spin polarized electrons and the local magnetic moments in the free layer~\cite{Slon,Berger}. The spin-torque induced precession of the free layer's magnetization has been experimentally verified in the presence of current and magnetic field up to 26 GHz~\cite{Kiselev,Mancoff1,Houss,Mancoff2} and around 18 GHz in magnetic tunnel junctions~\cite{Nazarov1,Nazarov2}.  The theoretical investigation on STNO predicts that the spin torque acts like an external power against damping of the magnetization and maintains self-oscillations. In the case of self-oscillations, the magnetization never converges to a final state as if the system had no damping at all.  With appropriate conditions between the spin torque, magnetic field and damping constant, it is possible to form a very stable precession even in the presence of large thermal fluctuations~\cite{Li}.

The STNOs are characterized by various parameters such as frequency, power, and Q-factor. Their efficient performance arises essentially due to their nano-scale dimension and ability to tune their oscillation frequencies with the enhancement of power and Q-factor.  Apart from the requirement of high frequency, tunability for a large range of frequencies by current or field is also desired for different applications~\cite{Z_Zeng}.  

Investigations on STNOs with perpendicularly magnetized free layer and parallelly magnetized pinned layer in the presence of an out-of-plane field show the possibility of (1) steady state precession with a maximum frequency of 6.3 GHz with a high power of 0.55 $\mu$W~\cite{Kubota} and Q-factor 135 and (2) steady state precession with a maximum frequency of 45 GHz with a high Q-factor over 3000 and power 64 nW~\cite{Maehara}. Without an out-of-plane field, the steady state precession is not possible for this configuration of STNO, and it can be induced by appropriate field-like torque~\cite{Taniguchi1,Guo}. 

A further enhancement of frequency in the STNOs with parallelly magnetized free layer has been achieved by applying an in-plane field~\cite{Bonetti,Suzuki1,Suzuki2,ZM_Zeng,Arun1} or by tilting the pinned layer's magnetization to out-of-plane~\cite{Zhou1,Zhou2,Zhou3,Zhou4,Lv,Arun2}.  In particular, Bonetti $et~al$. have experimentally observed the tunability of the frequency by current, magnetic field strength and magnetic field direction in nanocontact based STNOs~\cite{Bonetti}.  They have found frequencies up to 46 GHz for the strengths of magnetic field and current, ranging from 8 kOe to 14 kOe and from 10 mA to 20 mA, respectively.  Arun $et~al$., have studied the STNO in the presence of an in-plane magnetic field and found high frequency oscillations up to 68 GHz~\cite{Arun1}. Also, these authors have identified the oscillations with frequency up to 75 GHz in the absence of the field by tilting the polarization of the pinned layer to out-of-plane~\cite{Arun2}.  

Therefore, the high frequency self-oscillations up to or above 70 GHz in an STNO are achieved only by applying continuous in-plane/out-of-plane field~\cite{Bonetti,Arun1} or by tilting the polarization of the pinned layer~\cite{Arun2}. Especially, getting self-oscillations corresponding to the low energy initial magnetization state (i.e. aligned along easy axis direction) is practically a challenging task.  On the contrary, achieving the self-oscillations by current without applying continuous external field and tilting the polarization of the pinned layer still remains to be investigated for the case of parallelly magnetized free layer.   Therefore, in this present study, we examine the occurrence of self-oscillations in an STNO having parallelly magnetized free and pinned layers in the presence of a field-like torque.  We show the rapid response in the magnetization of the free layer towards the transition from a steady state (no oscillatory state) to a self-oscillatory state  after being triggered by the in-plane magnetic field applied for a short interval of time. The steady state is an equilibrium state where the magnetization converges to settle finally. We find that the  in-plane field applied for a short duration causes the onset of self-oscillations of the magnetization with frequency up to 72 GHz,  even in the absence of field-like torque. Also, we show the stability of the STNO against thermal noise and increment of Q-factor with frequency.

 The paper is organized as follows. Section II addresses the geometry of the STNO and the governing equation, namely the, Landau-Lifshitz-Gilbert- Slonczewski (LLGS) equation along with the details of the effective field. The occurrence of self-oscillations of the magnetization due to the in-plane field, the tunability of the frequency by current and field-like torque are presented in Section III. Finally, the conclusions are summarized in Section IV. We present the salient features of the derivation of the LLGS equation in Appendix A and then in Appendix B we provide the impact of the thermal noise on the oscillation frequency and power spectral density.

\section{Model Description and Governing Equation}
 \begin{figure}[htb]
 	\centering\includegraphics[angle=0,width=0.5\linewidth]{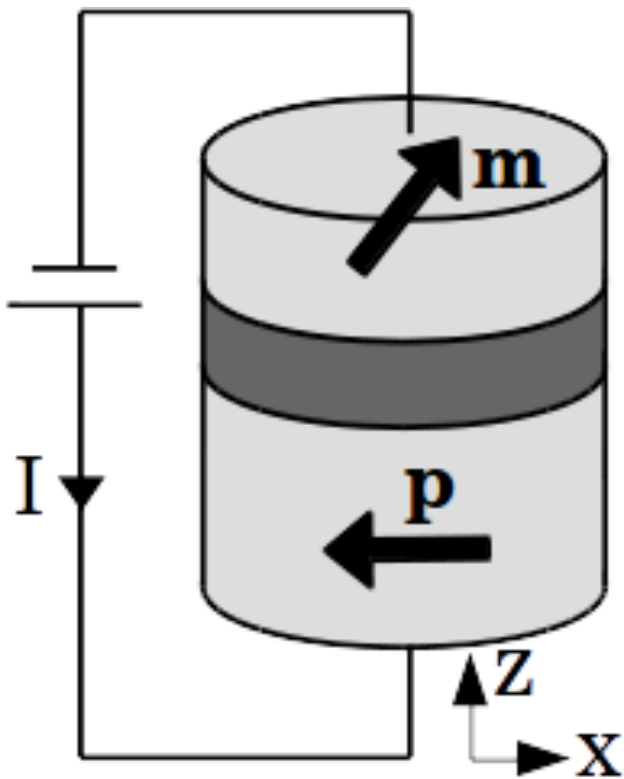}
 	\caption{Schematic picture of the model}
 	\label{model}
 \end{figure}
We consider an STNO as shown in Fig. \ref{model} with free and pinned layers (mild gray color), and a spacer (dark gray color).  The unit vectors of the magnetization of the pinned layer and free layer are denoted by ${\bf p}$ and ${\bf m}$, respectively. 

The LLGS equation that governs the dynamics of the magnetization of the free layer is given by (See Appendix A),
\begin{align}
\frac{d{\bf m}}{dt}=-\gamma ~ {\bf m}\times{\bf H}_{eff}+ \alpha~ {\bf m}\times\frac{d{\bf m}}{dt}+\gamma H_{S}~ {\bf m}\times ({\bf m}\times{\bf p})+\gamma\beta H_{S} ~{\bf m} \times{\bf p}. \label{llg}
\end{align}
In Eq. \eqref{llg},  ${\bf m} =  m_{x}~{\bf e}_x + m_{y}~{\bf e}_y + m_{z}~{\bf e}_z$ or ${\bf m} =(m_x,m_y,m_z)$, $|{\bf m}|$ = 1, where ${\bf e}_x$, ${\bf e}_y$ and ${\bf e}_z$ are the unit vectors along the positive $x$, $y$ and $z$ directions, respectively.  $\gamma$ is the gyromagnetic ratio, $\alpha$ is the Gilbert damping parameter, $H_{S}$ is the strength of the spin-transfer torque and  $\beta$ is the ratio between the strengths of field-like torque~\cite{Li,Zhang} and the spin-transfer torque.  The sign and magnitude of the field-like torque depend on the material parameters such as diffusion constant, spin-flip relaxation time, thickness of the free layer and the exchange interaction between the itinerant electron and the magnetic background~\cite{Zhang,Shpiro,Li1}.  $H_S$ is defined as
\begin{align}
H_{S} = \frac{H_{S0}}{1+\lambda~ {\bf m . p}},\label{H_S0}  
\end{align}
where $H_{S0}=\hbar\eta I/2 e M_s V$. Here $I$ is the current passing through the free layer,  $\hbar(=h/2\pi)$ is the reduced Plank's constant, $e$ is the electron charge, $M_s$ is the saturation magnetization of the free and pinned layers, $V$ is the volume of the free layer, $\eta$ and $\lambda$ are dimensionless parameters that determine the magnitude and the angular dependence of the spin-transfer torque, respectively. Here, the pinned layer's magnetization is fixed along the negative $x$-direction, and therefore ${\bf p}=-{\bf e}_x$.  The positive (negative) current corresponds to the flow of electrons from the free (pinned) to pinned (free) layer.  The effective field ${\bf H}_{eff}$ of the free layer is given by
\begin{align}
{\bf H}_{eff} ~=~& H_k m_{x}~{\bf e}_x-4\pi M_s m_{z}~ {\bf e}_z + H_a(t) ~{\bf n} \label{Heff},
\end{align}
where $H_k$ is the magneto-crystalline anisotropy field, $4\pi M_s$ is the demagnetization field and $H_a(t)$ is the externally applied time-dependent in-plane field, which can be applied continuously or for a short interval of time.   The latter case may be defined as a short in-plane field since it is applied for a few nanoseconds only and it triggers self-oscillations as we see in the next section.  The time up to which the short in-plane field is applied can be called cut-off time and it is denoted by $T_{off}$.  In Eq. \eqref{Heff}, ${\bf n}=\cos\phi_H ~{\bf e}_x+  \sin\phi_H~ {\bf e}_y$ is the unit vector along which the in-plane field $H_a$ is applied and  $\phi_H$ is the in-plane field angle between ${\bf n}$ and ${\bf e}_x$~\cite{Taniguchi2}. The effective field is defined by ${\bf H}_{eff} = - \partial E/\partial (M_s {\bf m})$, where the $E$ is the effective energy density given by~\cite{Taniguchi3}
\begin{align}
E =  &- \frac{M_s}{2}[ H_k ({\bf m}.{\bf e}_x)^2-  4\pi M_s ({\bf m}.{\bf e}_z)^2]-M_s H_a {\bf m}.{\bf n}.\label{E}
\end{align}

In our present study, we consider the situation where the pinned and free layers are made up of cobalt and the material parameters corresponding to cobalt are chosen as~\cite{Taniguchi2} $M_s$ = 1448.3 emu/c.c., $H_k$ = 18.6 kOe, $\eta$ = 0.54, $\lambda = \eta^2$, $\gamma$ = 17.64 Mrad/(Oe s), $\alpha$ = 0.005, $\mu_0$ = 1 and $V$ = $\pi \times 60 \times 60 \times 2$ nm$^3$. 

\section{Results}
\subsection{Oscillations triggered by in-plane magnetic field}
\begin{figure}[htb]
	\centering\includegraphics[angle=0,width=1\linewidth]{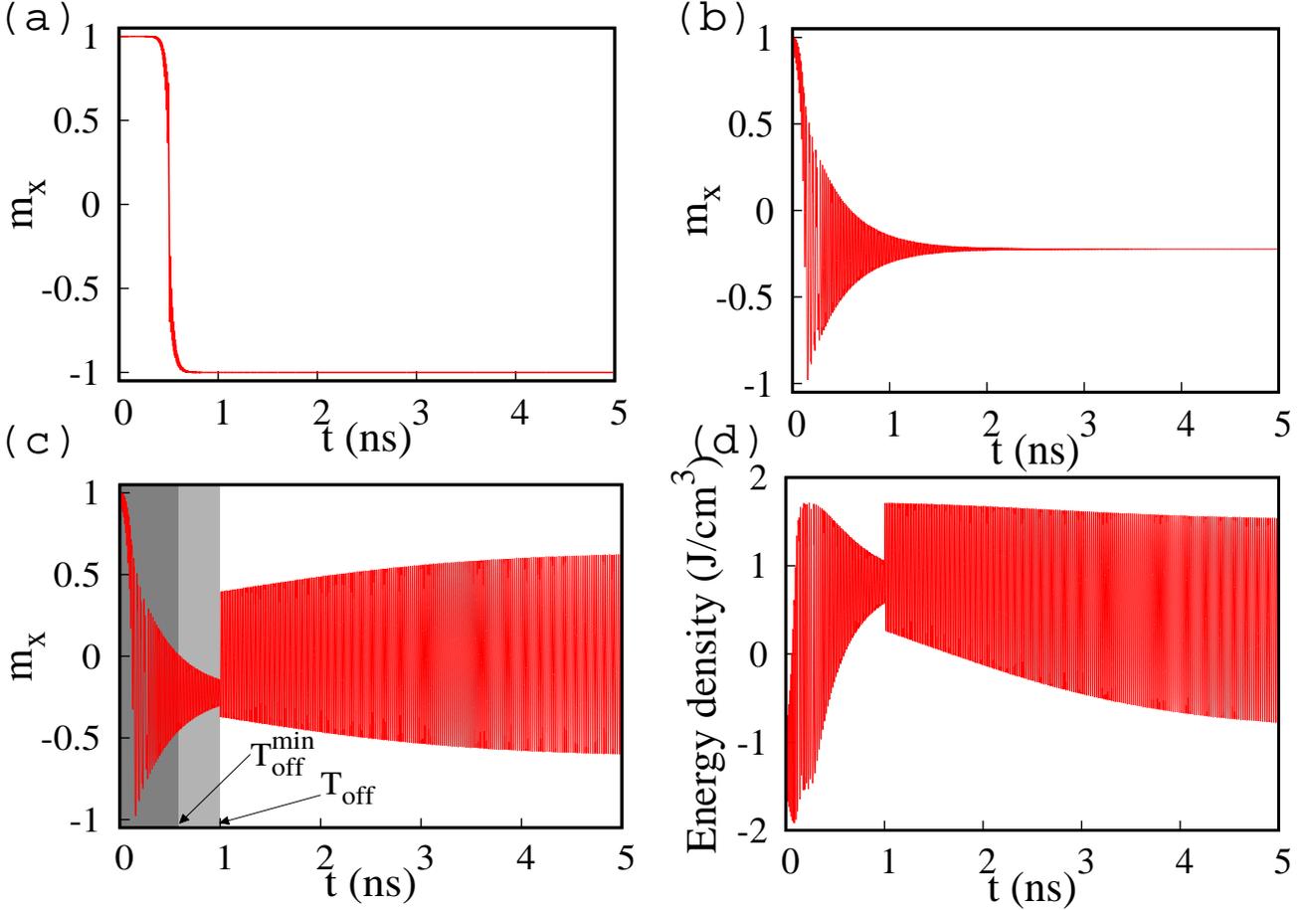}
	\caption{ Temporal evolution of $m_x$ (a) in the absence of in-plane field, (b) in the presence of in-plane field. Temporal evolution of (c) $m_x$ and (d) energy density in the presence of short in-plane field with the cut-off time $T_{off}$ = 1 ns, $H_a$ = 10 kOe and the direction $\phi_H$ = 30$^\circ$.  The mild and dark gray background colors in (c) indicate the cut-off time $T_{off}$ up to which the in-plane field is applied and the minimum cut-off time $T^{min}_{off}$, respectively. Here $\beta$ = 0.5 and $I$ = -15.5 mA.}
	\label{energy}
\end{figure}

\begin{figure}[htb]
	\centering\includegraphics[angle=0,width=1\linewidth]{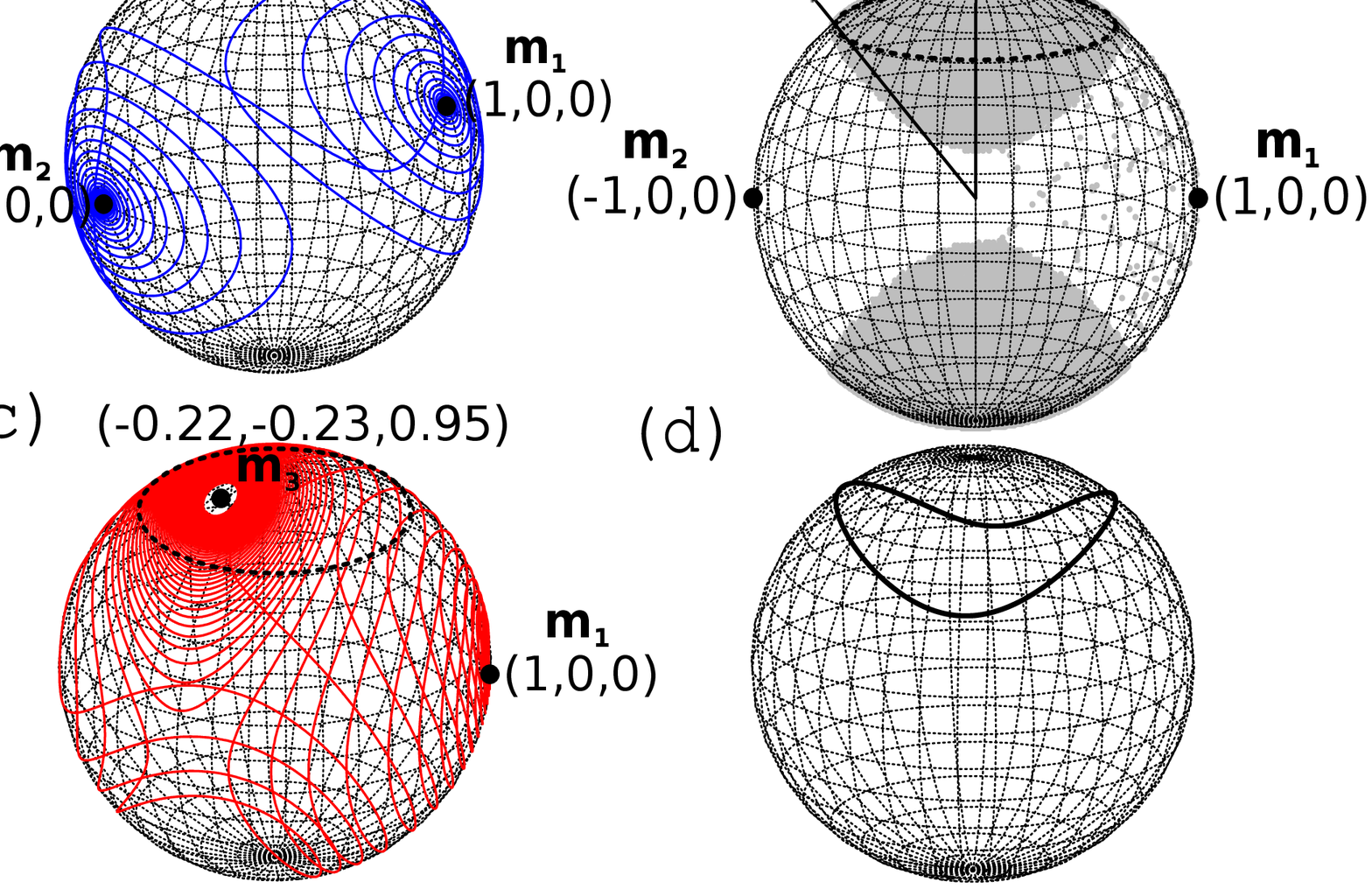}
	\caption{(a) Magnetization trajectories corresponding to the magnetization reversal from $m_x$ = +1 to -1 and (b) basins of self-oscillation state (gray points) and steady state  (white region) in the absence of field when $I$ = -15.5 mA and $\beta$ = 0.5. Magnetization trajectories of (c) the steady state due to the continuously applied in-plane field for $H_a$ = 10 kOe and $\phi_H$ = 30$^\circ$ and (d) a trajectory of the self-oscillation due to the short in-plane field with  $H_a$ = 10 kOe, $\phi_H$ = 30$^\circ$ and $T_{off}$ = 1 ns.  Here $\beta$ = 0.5. The dotted  black circles in (b) and (c) correspond to $\theta=\theta_{osc}$.}
	\label{trajectory}
\end{figure}

Equation \eqref{E} implies that there exist two low energy magnetization states ${\bf m}_1$ = (1, 0, 0) and ${\bf m}_2$ = (-1, 0, 0) (we call as primary equilibrium states) in the absence of current and in-plane field, marked by the black solid points in Figs.\ref{trajectory}(a) and \ref{trajectory}(b). Here, ${\bf m}_2$ is the linearly stable equilibrium state, and ${\bf m}_1$ is the unstable equilibrium point in the presence of current alone (i.e. in the presence of current and absence of the external field). It means that if we choose the initial magnetization state nearer to these states the system asymptotically reaches the equilibrium state ${\bf m}_2$ (Fig. \ref{energy}(a)). Though the state ${\bf m}_2$ is linearly stable it is nonlinearly unstable, which means that if we choose the initial magnetization state far away from ${\bf m}_1$ or ${\bf m}_2$ , the system settles into a new stable self-oscillatory state in the asymptotic limit. However, for $I\ne$0, $H_{a}\ne$0 and $\phi_{H}\ne$0, both ${\bf m}_1$ and ${\bf m}_2$ are no longer equilibrium states for the system.   In this case, the system exhibits a new equilibrium state ${\bf m}_{3}$ (and we call this as the secondary equilibrium state), which is stable and the value of the equilibrium point depends on the value of the current and applied field (See Fig. \ref{energy}(b)).  Interestingly, we observe that the system shows oscillations for the  initial magnetization state chosen nearer to ${\bf m}_1$ (or ${\bf m}_2$) by applying the external magnetic field for a short interval of time.

\begin{figure}
	\centering\includegraphics[angle=0,width=1\linewidth]{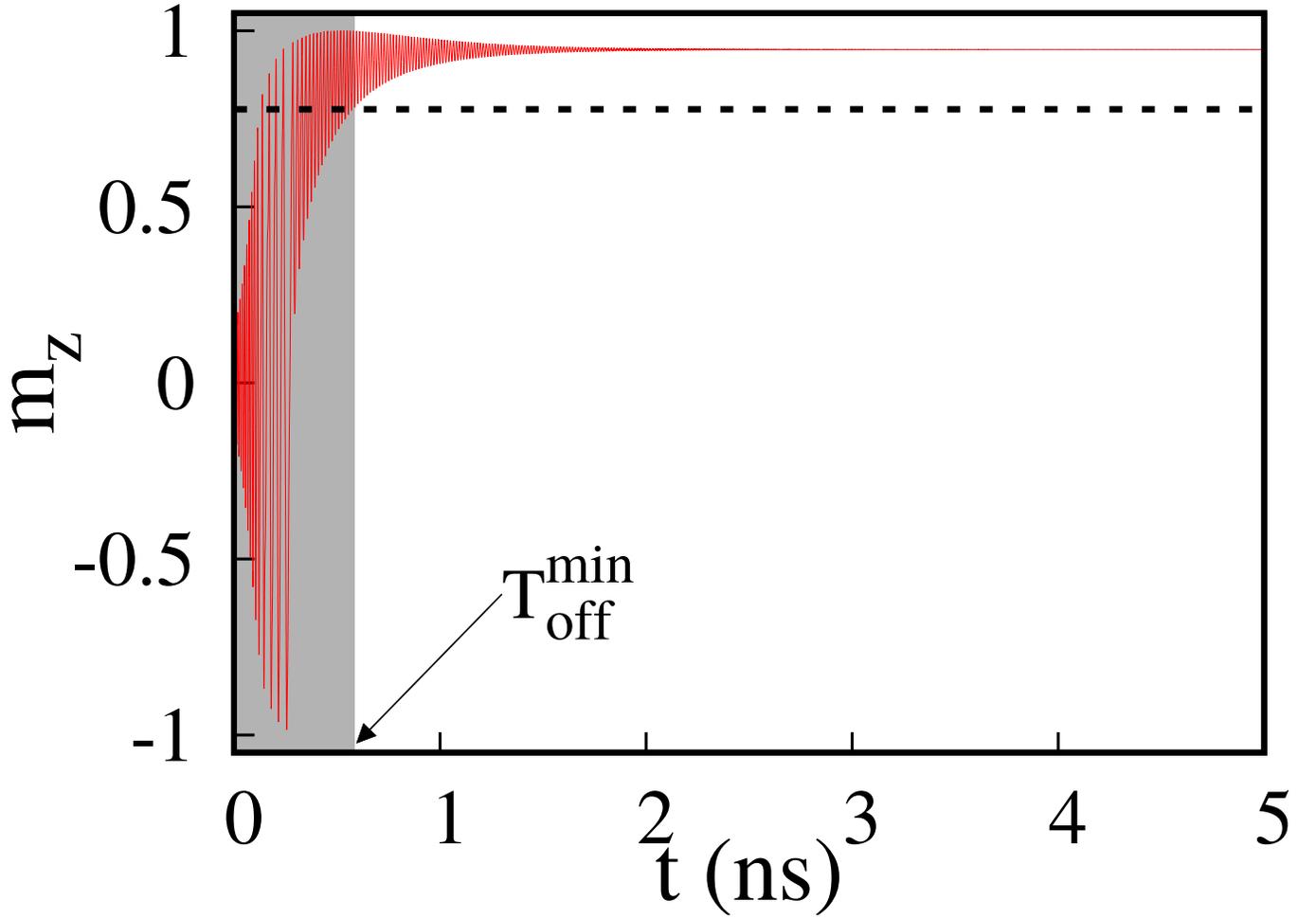}
	\caption{The temporal evolution of $m_z$ in the presence of continuous in-plane field.   The horizontal dashed line corresponds to $m_z = cos (\theta_{osc})$ for finding the minimum cut-off time $T_{off}^{min}$.  The gray background indicates the time up to $T_{off}^{min}$.}
	\label{T_min}
\end{figure}

In order to demonstrate the above, we fix the values of current the and the field-like torque as $I=-15.5$ mA, $\beta$=0.5 respectively, and examine the dynamics of the system in the absence and presence of the magnetic field in Figs.\ref{energy} and \ref{trajectory}.  For $H_{a}=0$, when the initial magnetization is chosen near the unstable fixed point ${\bf m}_1$, the system shows magnetization reversal and asymptotically reaches the stable state ${\bf m}_2$ as shown in Figs.\ref{energy}(a) and \ref{trajectory}(a).  Fig. \ref{trajectory}(b) shows the basin stability of the fixed point ${\bf m}_2$  as a white region, that is if the initial magnetization state is taken in this region, the magnetization reaches the state ${\bf m}_2$ asymptotically. If we choose the initial magnetization state in the gray region  which refers to the basin of self-oscillatory state in Fig. \ref{trajectory}(b),  the system will show self-oscillations of the magnetization. 
When we apply the in-plane field continuously with   $H_a$ = 10 kOe and, $\phi_{H}=30^{o}$, where the initial magnetization is again chosen near to the unstable fixed point ${\bf m}_1$, the system reaches a secondary equilibrium state ${\bf m}_3=(-0.22, -0.23,0.95)$ asymptotically, which is clearly evident from Fig. \ref{trajectory}(c). The corresponding time evolution of $m_{x}$ and $m_{z}$ are plotted in Figs.\ref{trajectory}(c) and \ref{T_min}, respectively.  

However, if we apply a short duration in-plane field, for example in the form

\begin{align}
{\bf H_{a}}(t)=
\begin{cases}
  10(\cos 30^{o} ~{\bf e}_{x}+ \sin 30^{o} ~{\bf e}_{y})  \text{ kOe}, & \text {  $t\leq T_{off}$} \\
 0, & \text {  $t>T_{off}$}  \nonumber
  \end{cases}  
\end{align}

 the system shows self-oscillations even when the initial magnetization is chosen in the basin of ${\bf m}_{2}$.
The magnetization trajectory corresponding to the self-oscillations of the magnetization  is plotted after leaving out the transients in Fig. \ref{trajectory}(d) for $T_{off}=1~ns$.

 The mechanism behind the emergence of self-oscillations of the magnetization is as follows. When we apply the in-plane field the system reaches the secondary equilibrium state and it emerges in the basin of self-oscillatory state. So, when we cut off the field the system shows self-oscillations even for the  initial magnetization state  chosen in the basin of ${\bf m}_{2}$.

Now, the temporal evolution of $m_z$ corresponding to the steady state approaching towards the equilibrium state ${\bf m}_3$, which is also plotted in Fig. \ref{T_min}, confirms that the magnetization settles into ${\bf m}_3$ after 2 ns. This implies that the system will exhibit self-oscillations on applying the short duration in-plane field with the cut-off time $T_{off}\geq$ 2 ns.   Since the self-oscillations is achieved only after $T_{off}$, reducing the cut-off time $T_{off}$ will reduce the delay in getting oscillations.  Therefore it is essential to find out the minimum cut-off time $T^{min}_{off}$, so that the self-oscillations can be achieved with $T_{off}\geq T^{min}_{off}$.

The $T^{min}_{off}$ is the time at which the magnetization enters the basin of  self-oscillatory state and continues to precess inside the basin, around the new steady state ${\bf m}_3$. One of the ways to estimate $T^{min}_{off}$ is by identifying the zenith angle $\theta_{osc}$ (see Fig. \ref{trajectory}(b)) in such a way that the region corresponding to 0$^\circ \leq \theta \leq \theta_{osc}$ does not include the basin of steady state. For instance, the dotted black circle corresponding to $\theta_{osc}$ = 39$^\circ$ drawn in Fig. \ref{trajectory}(b) does not include the basin of steady state. Here, $T^{min}_{off}$ can be considered as the time at which the magnetization enters the dotted circle and continues to precess inside the circle, around ${\bf m}_3$.  The time  at which the magnetization enters the dotted circle is the same as the time at which the $m_z$ begins to oscillate above the horizontal line drawn corresponding to $m_z = \cos\theta_{osc}$ (see Fig. \ref{T_min}). For instance, the minimum cut-off time $T^{min}_{off}$ is identified as 0.57 ns corresponding to the $\theta_{osc}$ = 39$^\circ$ as shown in Fig. \ref{T_min}. 

\begin{figure}
	\centering\includegraphics[angle=0,width=1\linewidth]{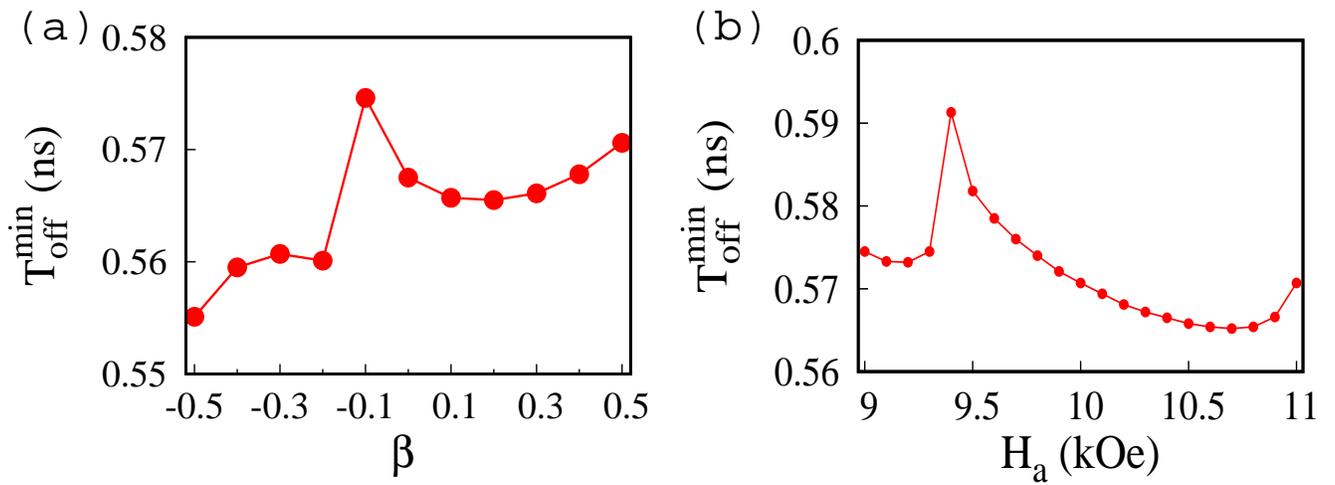}
	\caption{The minimum cut-off time $T_{off}^{min}$ against (a) $\beta$ for $H_a$ = 10 kOe and (b) $H_a$ for $\beta$ = 0.5. Here, $I$ = -15.5 mA.}
	\label{mincutoff}
\end{figure} 

The oscillations of the magnetization for $T_{off}$ = 1 ns  are confirmed in Fig. \ref{energy}(c), where $m_x$ is plotted for $I$ = -15.5 mA, $\beta$ = 0.5, $H_a (t)$ = 10 kOe($t\leq T_{off}$) and 0 Oe($t>T_{off}$), and $\phi_H$ = 30$^\circ$.  We can observe that the oscillations are damped out as $t\rightarrow T_{off}$ and triggered back as $t>T_{off}$ due to the short in-plane field.  By using the same procedure followed to determine the $T_{off}^{min}$ for $\beta$ = 0.5 we can also identify the values of $T_{off}^{min}$ for different values of $\beta$ and $H_a$ and these are plotted in Fig.\ref{mincutoff}(a) and Fig.\ref{mincutoff}(b), respectively.  Figs.\ref{mincutoff} show that the minimum cut-off time does not exceed $\sim$0.6 ns, which implies that the in-plane field need not be applied beyond 0.6 ns to trigger the self-oscillations. The slight change in $T_{off}^{min}$ due to $\beta$ and $H_a$ can be attributed to the fact that the size of the basin of self-oscillatory state does not change considerably due to the change in $\beta$ or $H_a$. 
 
The time evolution of the energy density during the process of oscillations due to the short in-plane field shown in Fig. \ref{energy}(c) is plotted in Fig. \ref{energy}(d). Fig. \ref{energy}(d) implies that in the presence of field and current the energy oscillates initially and then tends to  reach steady state energy of 0.8 J/cm$^3$. Consequently, the magnetization damps to reach the steady state ${\bf m}_3$ within the basin of the self-oscillatory state as shown in Figs.\ref{trajectory}(c) and \ref{T_min}.  After the in-plane field is switched off at time $T_{off}$(=1 ns) the energy density emerges back to steadily oscillate between 1.50 J/cm$^3$ and -0.88 J/cm$^3$ and therefore $m_x$ oscillates steadily after 1 ns as shown in Fig.\ref{energy}(c).

\begin{figure}
\centering\includegraphics[angle=0,width=1\linewidth]{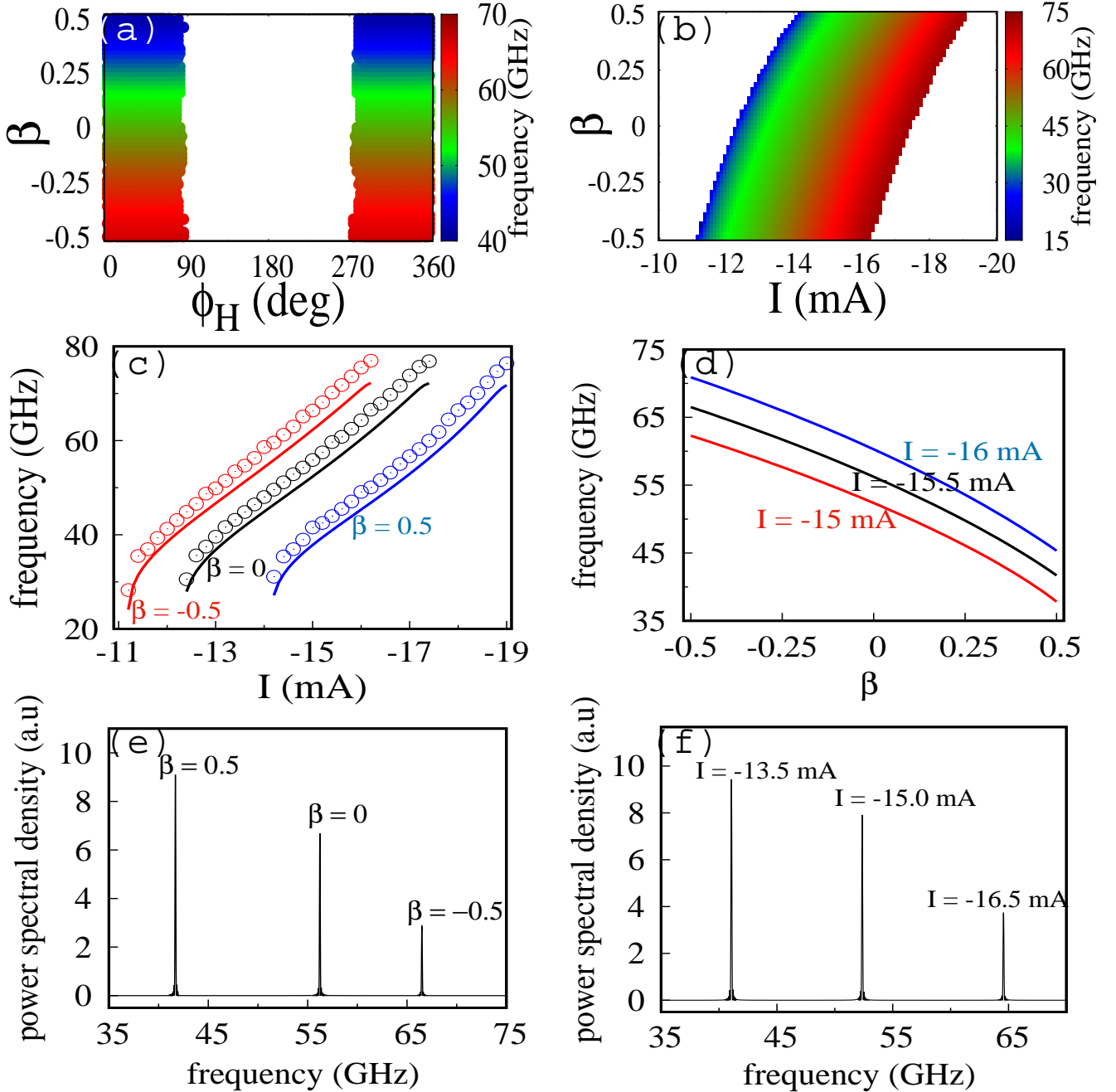}
\caption{Oscillation frequency with respect to (a) $\beta$ and $\phi_H$ when  $T_{off}$ = 1 ns and $I$ = -15.5 mA and (b) $\beta$ and $I$ when  $H_a$ = 10 kOe, $\phi_H$ = 30$^\circ$ and $T_{off}$ = 1 ns.  The numerically (solid lines) and analytically (open circles) obtained oscillation frequency in the presence of short in-plane field against (c) current and (d) \textcolor{blue}{$\beta$} for $H_a$ = 10 kOe, $\phi_H$ = 30$^\circ$ and $T_{off}$ = 1 ns. Power spectral density in the presence of short in-plane field for different values of (e) $\beta$ while $I$ = -15.5 mA and (f) current while $\beta$ = 0 for $\phi_H$ = 30$^\circ$.  Here, $T_{off}$ = 1 ns.}
\label{power}
\end{figure}

\subsection{Frequency and power spectral density}

We also find that the self-oscillations of the magnetization are not triggered for all the directions of the short duration in-plane field. This is because the position of the steady state ${\bf m}_3$ formed by the in-plane field depends on the direction of the short duration in-plane field.  For some directions of the short duration in-plane field, the state ${\bf m}_3$ is formed inside the basin of the steady state  and, consequently, no oscillation is exhibited. In order to verify the range of $\phi_H$ for which the self-oscillation is possible, we plot the oscillation frequency between $\beta$ and $\phi_H$ in Fig. \ref{power}(a) for  $T_{off}$ = 1 ns and $I$ = -15.5 mA.  The values of ($\beta,\phi_H$) corresponding to the white region between $\phi_H$ = 90$^\circ$ and 270$^\circ$ exhibit self-oscillations and the magnetization settles into ${\bf m}_2$.  This is because when $\phi_H$ is between 90$^\circ$ and 270$^\circ$, ${\bf m}_3$ is formed outside the basin of self-oscillatory state and therefore no self-oscillatory state is identified in the presence of the short in-plane field.  The system exhibits oscillations for $\phi_H$ between 0$^\circ$ and 90$^\circ$ and 270$^\circ$ and 360$^\circ$, where it is observed that the frequency is independent of the in-plane field angle $\phi_H$.  

Similarly, the current dependence of the self-oscillations is also identified by plotting the frequency against $\beta$ and $I$ for $\phi_H$ = 30$^\circ$ and $T_{off}$ = 1 ns in Fig. \ref{power}(b) and it is observed that the oscillatory region lies between the nonoscillatory regions (white regions). From Fig. \ref{power}(b) we can see that the frequency is minimum at $I = I_c^{low}$ and maximum at $I=I_c^{upp}$.  For $\beta=0$, $I_c^{low}$ is -12.4 mA and  $I_c^{upp}$ is -17.3 mA. Thus we conclude that self-oscillations are possible only when the magnitude of the current is above the lower critical current ($I_c^{low}$) and below the upper critical current ($I_c^{upp}$).  The values of both the critical currents decrease when the field-like torque is negative ($\beta<0$). Further, from Fig. \ref{power}(b) we can also observe an enhancement in the frequency as a function of the current and field-like torque. The frequency spectra for different values of field-like torque and current are plotted in Fig. \ref{power}(c) and (d) against the current and \textcolor{blue}{$\beta$}, respectively. In Fig. \ref{power}(c) we have plotted the numerically computed frequency against the current by solid lines for different values of $\beta$ = 0, 0.5 and -0.5 in the presence of the short duration in-plane field with $\phi_H$ = 30$^\circ$ and $T_{off}$ = 1 ns.  From this figure, we can confirm a large range of tunability in frequency from $\sim$25 GHz to $\sim$72 GHz by the current, and this is important for applications for producing a wide range of frequencies.   The open circles correspond to the frequencies obtained from the expression 

\begin{align}
&f = \frac{\gamma\widetilde{m}_z}{2\pi (1+\alpha^2)}\nonumber\\&\left[ -\frac{H_k}{2}-4\pi M_s + \frac{H_{S0}(\alpha-\beta)}{\lambda (1-\widetilde{m}_z^2)}\left( \frac{1}{\sqrt{1-\lambda^2 (1-\widetilde{m}_z^2)}}-1\right)\right], \label{freq}
\end{align}

solved with the approximation $dm_z/dt=0$  by the procedure followed in Ref.[31].  In the expression \eqref{freq}, $\widetilde{m}_z$ is the average value of $m_z$ computed over $n$ number of precessions after reaching the self-oscillations. The small deviation between the analytical and numerical frequencies may be attributed to the approximation $m_z(t) = constant$ used in this analysis. Fig. \ref{power}(c) implies that the self-oscillations of the magnetization can be triggered by the short duration in-plane field even in the absence of field-like torque and that $\beta < 0$ is advantageous over $\beta > 0$ for getting the high frequency oscillations. To confirm the occurrence of self-oscillations for the appropriate range of $\beta$ (from -0.5 to 0.5, for Cobalt), we have plotted the graph of frequency against $\beta$ for different values of current, namely $I$ = -15 mA, -15.5 mA and -16 mA, in Fig. \ref{power}(d) for the short in-plane field with  $\phi_H$ = 30$^\circ$ and $T_{off}$ = 1 ns. As we can see in Fig. \ref{power}(d), for a given current, the field-like torque can increase the frequency up to 10 GHz.

The Q-factor (ratio between the peak frequency and line-width) of the STNO is expected to be large for the betterment of its applications.  In order to determine the Q-factor for different frequencies, we have plotted the power spectral density corresponding to different values of field-like torque for $I$ = -15.5 mA and current for $\beta = 0$ in Figs.\ref{power}(e) and (f), respectively, in the presence of the short in-plane field with  $\phi_H$ = 30$^\circ$ and $T_{off}$ = 1 ns. From the power spectral density plotted in Figs.\ref{power}(e), the Q-factor for the peaks corresponding to $\beta$ = 0, 0.5 and -0.5 are estimated to be 471.62, 636.82 and 753.11, respectively.  Similarly, from Fig. \ref{power}(f), the Q-factor for the peaks corresponding to $I$ = -13.5 mA, -15.0 mA and -16.5 mA are given by 463.92, 591.63 and 743.42, respectively.  The estimated values of the Q-factor imply that it gets enhanced considerably with the enhancement of the frequency.  Also in Appendix B the impact of the thermal noise on the oscillation frequencies and  the power spectral density is discussed.

The role of the in-plane magneto-crystalline anisotropy in the free layer is important for the oscillations triggered by the short in-plane field. We have verified that when the magneto-crystalline anisotropy is along the positive z-direction the short-inplane field triggers no magnetization oscillations and the magnetization only settles at (1,0,0) for positive current and at (-1,0,0) for negative current irrespective of field-like torque. The free layer with out-of-plane magneto-crystalline anisotropy exhibits the possibility of oscillations with a frequency range of $\sim$6 GHz only in the presence of a continuous perpendicular field~\cite{Arun1}. 

\subsection{Impact of free layer volume on critical currents}

\begin{figure}
	\centering\includegraphics[angle=0,width=1\linewidth]{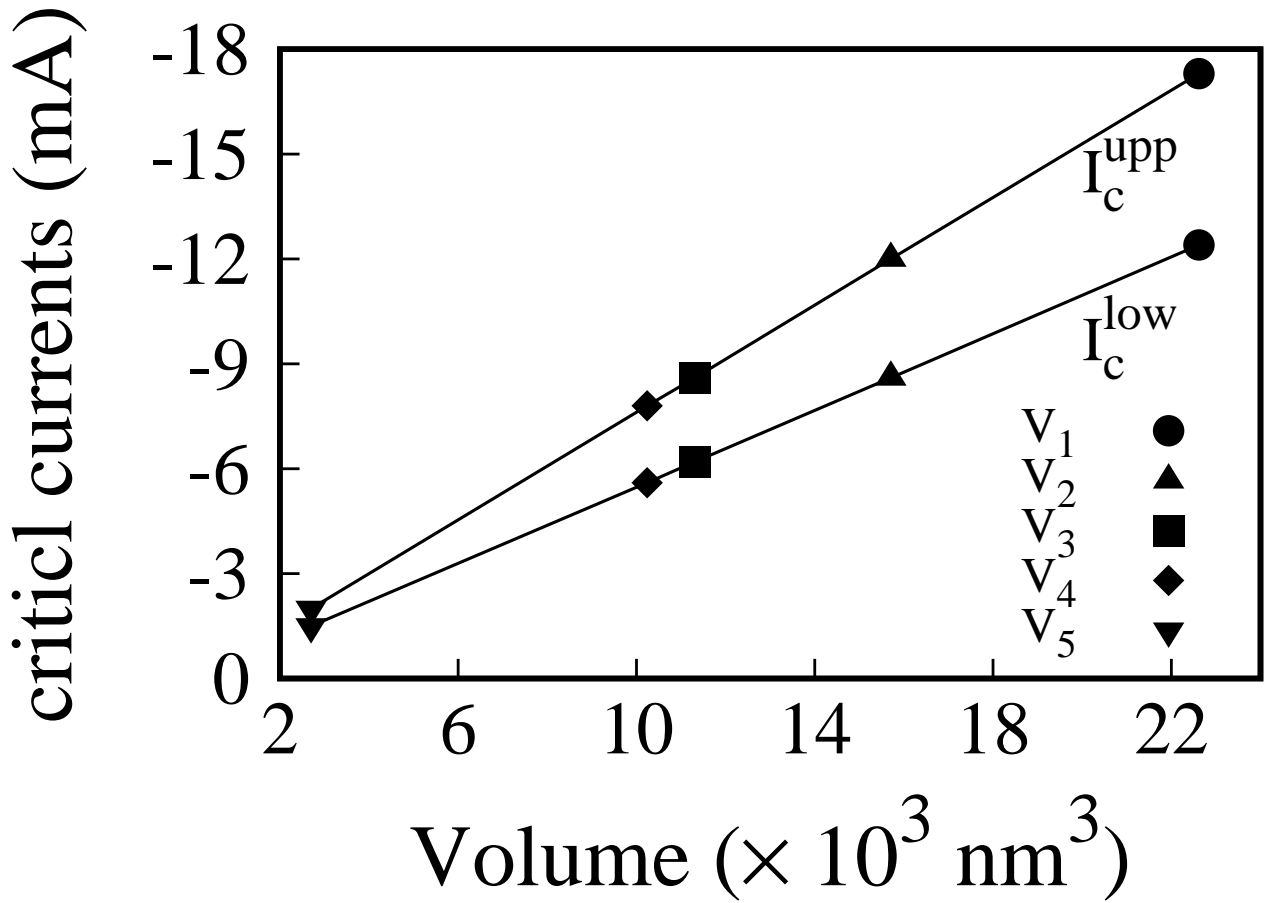}
	\caption{Critical currents for the onset and offset of the steady state precession for different volumes of the free layer in the presence of the short  duration in-plane field. Here  $H_a$ = 10 kOe, $\phi_H$ = 30$^\circ$, $T_{off}$ = 1 ns and $\beta$ = 0.}
	\label{critI_vs_vol}
\end{figure}

The theory discussed above focuses on the spin-torque oscillator dynamics of the magnetization vector ${\bf m}$, corresponding to the volume of the free layer $V$ = 2$\times \pi\times$60$\times$60 nm$^3$.  We have found that the magnitude of the critical currents $I_c^{low}$ and $I_c^{upp}$ are 12.4 mA and 17.3 mA, respectively, in the presence of the short duration in-plane field with $\phi_H$ = 30$^\circ$ and $T_{off}$ = 1 ns for $\beta$ = 0. The above range of current (above 10 mA) and field (10 kOe) may be considered to be quite high but we wish to note here that such values have been operated for STNO previously. Bonetti $et~al$., have experimentally observed such oscillations in STNO with the in-plane field 14.5 kOe and current 20 mA~\cite{Bonetti}.  Also, Taniguchi $et~al$., have theoretically studied for the relaxation time in STNO for the range of current 0 to 25 mA~\cite{Taniguchi2}. Nevertheless, the magnitudes of the critical currents can be decreased further by decreasing the volume of the free layer.  To confirm that this is indeed true, we have plotted the critical currents $I_c^{low}$ and $I_c^{upp}$  for different volumes of the free layer, namely $V_1$ = 2$\times \pi\times$60$\times$60 nm$^3$~~\cite{Arun1,Arun2,Taniguchi2}, $V_2$ = 2$\times \pi\times$50$\times$50 nm$^3$, $V_3$ = 1$\times \pi\times$60$\times$60 nm$^3$, $V_4$ = 2.5$\times$64$\times$64 nm$^3$~~\cite{Li} and $V_5$ = 3$\times$30$\times$30 nm$^3$~~\cite{Sbiaa}.   The results are shown in Fig. \ref{critI_vs_vol} which essentially imply that the critical currents at which the system exhibits self-oscillations of the magnetization that are directly proportional to the volume of the free layer.  Thus, a  decrease in the volume of the free layer decreases the critical currents required for the onset and offset of the self-oscillations.

\section{Conclusions}

We have studied an STNO with parallelly magnetized free and pinned layers in the presence of field-like torque by solving the LLGS equation. We have systematically shown the methodology to obtain self-oscillations of the free layer's magnetization by applying an in-plane field for a short duration of time ($<$1ns) corresponding to the low energy initial magnetization state. Expression for the frequency was derived and its validity was confirmed by numerical simulation. We may emphasize that the self-oscillations of the magnetization manifest even in the absence of field-like torque and one can realize the possibility of enhancement in the oscillation frequency up to 10 GHz in the presence of field-like torque.  We have also explored the tunability of the frequency from $\sim$25 GHz to $\sim$72 GHz by the current.  We determined the Q-factor for different frequencies and confirmed that the Q-factor increases with frequency. Finally, we have also confirmed that thermal noise does not affect the dynamics and the oscillation frequency appreciably.

\section*{Acknowledgements}
The works of V.K.C. and R. G are supported by the DST-SERB-CRG Grant No. CRG/2020/004353 and they wish to thank DST, New Delhi for computational facilities under the DST-FIST programme (SR/FST/PS-1/2020/135) to the Department of Physics.   M.L. wishes to thank the Department of Science and Technology for the award of DST-SERB National Science Chair.

\section*{Appendix A: Landau-Lifshitz-Gilbert-Slonczewski equation}
In this Appendix, we present some brief details on the derivation of the LLGS equation. For more details, we may refer to \cite{Lakshman}. The magnetic moment of an electron due to its orbital motion ${\boldsymbol{ \mathcal{ M}}_{orb}}$ in a circular orbit is given by
\begin{align}
{\boldsymbol{ \mathcal{ M}}_{orb}} = I_e {\bf A}, \tag{A.1} \label{Morb}
\end{align}
where $I_e$ is the current due to the orbiting electron and ${\bf A}$ is the vector area of the orbital in which the electron is revolving.  For an electron, ${\bf A} = -\pi r^2 \hat{L}$, where $r$ is radius of the circular orbit and $\hat{L}$ is the unit vector of orbital angular momentum of the electron.  If the electron moves with a uniform speed $v$ in the circular orbit with a time period $T$, we can write $I_e$ as, $I_e = e/T ={e}/{(2\pi r/v)}$. By substituting $I_e$ and ${\bf A}$ in Eq.\eqref{Morb}, we can get
\begin{align}
{\boldsymbol{ \mathcal{ M}}_{orb}} = -\frac{evr}{2} ~\hat{L}. \tag{A.2} \label{Morb1}
\end{align}
For the circular orbit motion, the orbital angular momentum of the electron {\bf L} is given by ${\bf L} = rmv~\hat{L}$, where $m$ is mass of the electron. Now, Eq.\eqref{Morb1} can be written as ${\boldsymbol{ \mathcal{ M}}_{orb}} = -({\mu_B}/{\hbar}){\bf L}$, where ${\mu}_B = e\hbar/2m$ is the Bohr magnetron. Analogously, one can define the magnetic moment due to the spin of an electron as
\begin{align}
{\boldsymbol{ \mathcal{ M}}_{spin}} = -\gamma'~{\bf S}, \tag{A.3} \label{Mspin}
\end{align}
where $\gamma'=g\frac{\mu_B}{\hbar}$. $\gamma'$ is the gyromagnetic ratio, $g$ is the g-factor and ${\bf S}$ is the spin angular momentum. The torque ${\bf N}$ exerted on the spin magnetic moment which is immersed in the magnetic field $\boldsymbol{{H}}$ is given by
\begin{align}
{\bf N} = \frac{d{\bf S}}{dt} = \mu_0~ {\boldsymbol{ \mathcal{ M}}_{spin}}\times {{\bf H}}, \tag{A.4} \label{N}
\end{align}
where $\mu_0$ is permeability in vacuum. By substituting ${\bf S}$ from Eq.\eqref{Mspin} in Eq.\eqref{N} we can derive
\begin{align}
\frac{d{\boldsymbol{ \mathcal{ M}}_{spin}}}{dt} = -\gamma''~ {\boldsymbol{ \mathcal{ M}}_{spin}}\times {{\bf H}}, \tag{A.5} \label{llgs1}
\end{align}
where $\gamma'' = \gamma' \mu_0$.  For each spin magnetic moment within a small volume $dV$ the Eq.\eqref{llgs1} can be written as
\begin{align}
\frac{d{\boldsymbol{ \mathcal{ M}}_{spin,i}}}{dt} = -\gamma''~ {\boldsymbol{ \mathcal{ M}}_{spin,i}}\times {{\bf H}}. \tag{A.6} \label{llgs2}
\end{align}
Here the small volume $dV$ is chosen in such a way that the magnetic field is uniform inside it. If we take volume average on both sides of Eq.\eqref{llgs2}
\begin{align}
\frac{d{ \left(\sum_i\boldsymbol{\mathcal{ M}}_{spin,i}/dV\right)}}{dt} = -\gamma''~ {\left(\sum_i\boldsymbol{\mathcal{ M}}_{spin,i}/dV\right)}\times {\bf{H}}.\tag{A.7}\label{llgs3}
\end{align} 
The quantity $\sum_i\boldsymbol{\mathcal{ M}}_{spin,i}/dV$ is the net magnetic moment per unit volume and it is denoted as the magnetization ${\bf M}$. Then Eq.\eqref{llgs3} can be rewritten as
\begin{align}
\frac{d\bf M}{dt} = -\gamma'' ~ {\bf M}\times{\bf H}. \tag{A.8}\label{llgs4}
\end{align}
Here ${\bf M} = M_s{\bf m}$, where $M_s$ is the magnitude of the magnetization called saturation magnetization and ${\bf m}$ is the unit magnetization vector.  Equation \eqref{llgs4} represents the steady precession of ${\bf M}$ around the field ${\boldsymbol{ H}}$. But in reality the magnetization will damp and align with the field after finite time due to energy dissipation. This damping can be represented phenomenologically by introducing an additional term in Eq.\eqref{llgs4} as \cite{Ralph,Landau,Gilbert}
\begin{align}
\frac{d\bf M}{dt} = -\gamma'' ~ {\bf M}\times{\bf H} - \frac{\lambda}{M_s}~{\bf M}\times\left({\bf M}\times{\bf H} \right). \tag{A.9}\label{llgs5}
\end{align}
Equation \eqref{llgs5} is called Landau-Lifshitz equation was first derived by Landau and Lifshitz in 1935 to understand the dynamics of the magnetization in the presence of magnetic field for small damping~\cite{Landau,Lakshman}. The damping term in Eq.\eqref{llgs5} was replaced with a more convincing form  for large damping in 1955 by Gilbert~\cite{Gilbert} and the modified Landau-Lifshitz equation is given by
\begin{align}
\frac{d\bf M}{dt} = -\gamma~{\bf M}\times {\bf H} + \frac{\alpha}{M_s}~ {\bf M}\times\frac{d{\bf M}}{dt}. \tag{A.10}\label{llgs6}
\end{align} 
The two forms Eq.\eqref{llgs5} and Eq.\eqref{llgs6} are known to be equivalent with $\gamma''=\gamma/(1+\alpha^2)$ and $\lambda=\gamma\alpha/(1+\alpha^2)$.  After dividing by $M_s$, Eq. \eqref{llgs6} is rewritten for the evolution of ${\bf m}$ as
\begin{align}
\frac{d\bf m}{dt} = -\gamma~{\bf m}\times {\bf H} + \alpha~ {\bf m}\times\frac{d{\bf m}}{dt}. \tag{A.11}\label{llgs7}
\end{align}
Eq. \eqref{llgs6} or Eq. \eqref{llgs7} is now called the Landau-Lifshitz-Gilbert equation and the magnetic field ${\bf H}$ is now replaced by the ${\bf H}_{eff}$ which includes the magneto-crystalline anisotropy field, demagnetization field and external field for a uniformly magnetized material. Later, in order to study the influence of current polarized along with the direction ${\bf p}$ on the magnetization an in-plane spin-transfer torque term $\gamma H_{S}~ {\bf m}\times ({\bf m}\times{\bf p})$ was introduced by Slonczewski in 1996~\cite{Slon}. Then to investigate the impact of a perpendicular spin-transfer torque  the term  $\gamma\beta H_{S} ~{\bf m} \times{\bf p}$ was introduced by Zhang $et~al.,$ in 2002~\cite{Zhang,Li1}. With all these torques, the Landau-Lifshitz-Gilbert-Slonczewski equation is given by
\begin{align}
\frac{d{\bf m}}{dt}=-\gamma ~ {\bf m}\times{\bf H}_{eff}+ \alpha~ {\bf m}\times\frac{d{\bf m}}{dt}+\gamma H_{S}~ {\bf m}\times ({\bf m}\times{\bf p})+\gamma\beta H_{S} ~{\bf m} \times{\bf p}. \tag{A.12}\label{llgs8}
\end{align}
This is the starting equation \eqref{llg} for the unit vector of the magnetization.

\section*{Appendix B: Impact of Thermal Noise}
\begin{figure}
	\centering\includegraphics[angle=0,width=1.1\linewidth]{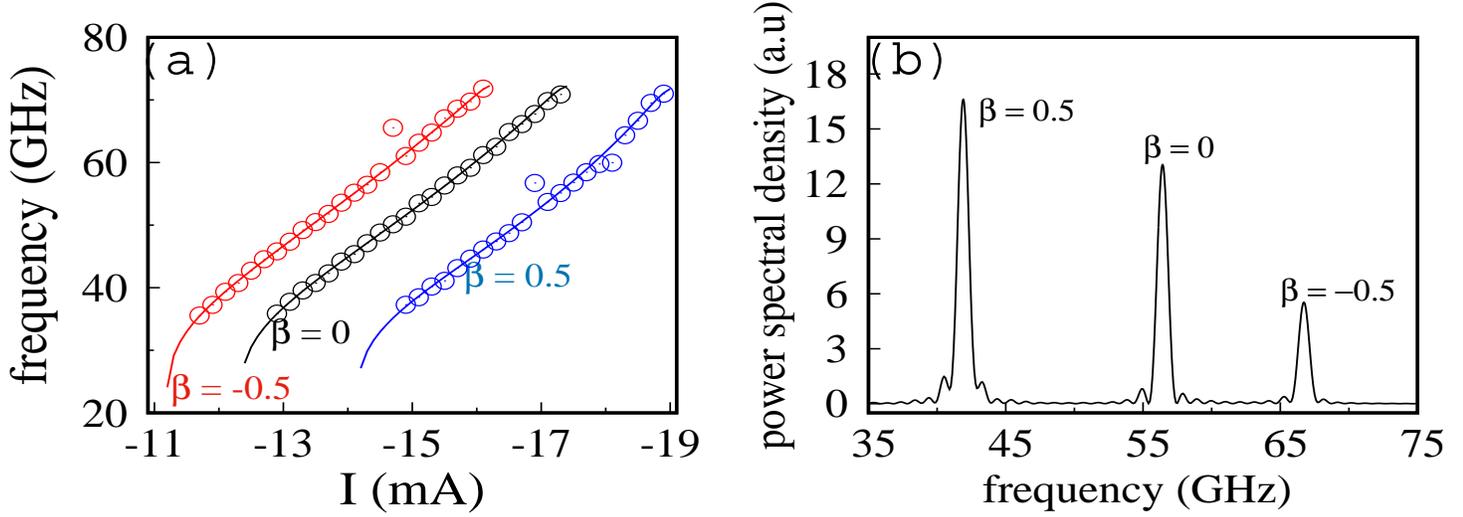}
	\caption{(a) Oscillation frequency due to short duration in-plane field corresponding to the temperatures $T$ = 0 K (solid line) and $T$ = 300 K (open circle). (b) Power spectral density in the presence of thermal noise corresponding to $T$ = 300 K and $I$ = -15.5 mA.  Here $T_{off}$ = 1 ns, $H_a$ = 10 kOe and $\phi_H$ = 30$^\circ$.}
	\label{noise}
\end{figure}
For practical applications, the dynamics of an STNO are required to be stable against the temperature.  Therefore, it is important to verify the impact of the thermal noise on the frequency of the magnetization oscillations and this is studied by including the thermal field due to the thermal noise in the effective field as follows:
\begin{align}
{\bf H}_{eff} ~=~& H_k m_{x}~{\bf e}_x-4\pi M_s m_{z}~ {\bf e}_z \nonumber\\
&+H_a(t) (\cos\phi_H ~{\bf e}_x+  \sin\phi_H~ {\bf e}_y)+ {\bf H}_{th}, \tag{B.1}
\end{align}
where the thermal field is given by
\begin{align}
{\bf H}_{th} = \sqrt{D}~ {\bf G},~~~~~D = \frac{2\alpha k_B T}{\gamma M_s \mu_0 V \triangle t}. \tag{B.2}
\end{align}
In the above, ${\bf G}$ is the Gaussian random number generator vector of the oscillator with components $(G_{x} , G_{y} , G_{z} )$, which satisfies the statistical properties $< G_{m} (t) >= 0$ and $< G_{m} (t) G_{n} (t') >= \delta_{mn}\delta(t-t')$ for all $m, n = x, y, z$.  $k_B$ is the Boltzmann constant, $T$ is the temperature, $\mu_0$ is the magnetic permeability in free space and $\triangle t$ is the step size of the time scale used in the simulation.  

To figure out the impact of the thermal noise on the frequency and power spectral density, Figs.\ref{noise}(a) and (b) are plotted  for different values of field-like torques in the presence of short duration in-plane field corresponding to $T_{off}$ = 1 ns and $\phi_H$ = 30$^\circ$ for the temperatures $T$ = 0 K and $T$ = 300 K.  In Fig.\ref{noise}(a), the frequencies corresponding to  $T$ = 0 K and 300 K have been represented by the solid lines and the open circles, respectively.  From Fig.\ref{noise}(a) and by comparing Figs.\ref{noise}(b) and \ref{power}(e), it can be understood that the results are not altered by the thermal noise in any perceptible way.

\end{document}